# DISCOVERY OF X-RAY CYCLOTRON ABSORPTION LINES MEASURES THE MAGNETIC FIELD OF AN ISOLATED NEUTRON STAR

G.F. Bignami[*,+], P.A. Caraveo[$], A. De Luca [$,#], S. Mereghetti[$]

\* Centre d'Etude Spatiale des Rayonnements, CNRS-UPS,  9 avenue du Colonel Roche ,
 31028 Toulouse Cedex4, France

+ Università degli Studi di Pavia, Dip. Fisica Nucleare e Teorica, Via Bassi, 6, 27100 Pavia - Italy

$ Istituto di Astrofisica Spaziale e Fisica Cosmica, Sezione di Milano "G. Occhialini",
   Via Bassini, 15, 20133 Milano – Italy

# Università di Milano Bicocca, Dip. Fisica, P.za della Scienza, 3, 20126 Milano - Italy

Isolated neutron stars are highly magnetized, fast-rotating end points of stellar evolution. They are now becoming directly observable through X-ray astronomy, owing to their high surface temperatures. In particular, features in their X-ray spectra could reveal the presence of atmospheres [1], or gauge their unknown magnetic fields through the cyclotron process [2], as in the classic case of X-ray binaries [3,4]. All isolated neutron stars spectra observed so far, however, appear as featureless thermal continua [5,6]. The unique exception is 1E1207.4-5209 [7,8,9]. In its spectrum, previous observations [10,11] had detected two deep absorption features, still too undefined for any unambiguous interpretation. Here we report on a much longer X-ray observation, in which the star's spectrum shows three distinct features, regularly spaced at .7, 1.4 and 2.1 keV, plus a fourth, lower significance one, at 2.8 keV. Such features vary in phase with the star rotation. Cyclotron resonant absorption is their logical interpretation, yielding a magnetic field strength of $8 \times 10^{10}$ G in the case of electrons. This is the first direct measure of an isolated neutron star magnetic field.

In August 2002, XMM-Newton devoted two orbits to 1E1207.4-5209 for a total observing time of 257,303 sec, or the longest observation with the EPIC instrument [12,13] on a galactic source. Selecting photons with energies between 0.2-4 keV, inside a 45" radius from the source, 208,000 photons were found in the pn [12] and 74,600 and 76,700, respectively, in the mos1 and 2 [13]. Figure 1 shows the spectra of the source as measured by the pn (figure 1a) and the two mos detectors summed together (figure 1b). Both independent spectral distributions show, in addition to the two wide absorption features centred at 0.7 and 1.4 keV [10,11], a third feature around 2.1 keV. The latter feature is seen in the three independent pn and mos1-mos2 data sets and was marginally visible also in the shorter XMM-Newton observation of December 2001 [11]. The statistical significance of the third feature can now be gauged by the result of an F-test. Its inclusion in the overall spectral fit (see caption) as a gaussian line in absorption yields an improvement in the quality of the fit which has a chance occurrence probability of $\sim 2\ 10^{-5}$ for the pn and $\sim 10^{-3}$ for each of the mos. A fourth dip at $\sim$2.8 keV is also seen in the pn data and its chance occurrence probability is estimated to be $\sim 5\ 10^{-2}$.

We are aware of the existence of edges of instrumental origin due to Au M (2.209 keV) and Si K (1.839 keV) in the general region of our feature at 2.1 keV. In the spectra of very bright sources, such edges can produce small calibration leftovers (e.g. [14]), seen at most at a level of 5%, fully compatible with the EPIC calibration accuracy [15], with an equivalent width < 10 eV (S.Molendi private communication). The ~2.1 keV feature present in our spectra is much stronger (with a ~15 times higher equivalent width and deviations from the continuum model at the level of ~25% as opposed to <5%). Moreover, the FWHM of our feature is 370 eV, as opposed to the value of 90 eV measured for the calibration leftovers. These characteristics give us confidence about the reality and the non-instrumental nature of our third feature.

The three absorption features shown in Figure 1 are wider than the CCD energy resolution and their intensities are definitely different. No significant substructure was detected in any of them. The shapes of the two main features cannot be described by any single gaussian profile. While detailed analytical or numerical fits to the feature shapes are beyond the scope of this work, we note that at least two gaussian profiles, always larger than the instrument resolution, are needed to describe each of the features at 0.7 and 1.4 keV (see also caption of Fig.1). In any case, Doppler broadening by thermal electron motion in the kT~ 0.2 keV environment is smaller than the CCD energy resolution. Unfortunately, the higher spectral resolution of the RGS [16] cannot provide any additional clue on this source since its limited sensitivity requires a drastic rebinning of the dispersed data. The resulting spectrum is coarser than those shown in figure 1.

After the first EPIC observation, variations of the lines as a function of the pulsar phase were reported [11]. In order to investigate such an effect in our longer dataset, we had first to search for the best source period. This was done following the steps outlined by Mereghetti et al [11] and yielded a P= 0.42413076 ± 0.00000002 s . The light curve for the total energy range (0.2-4 keV), shown in figure 2, is characterized by a nearly sinusoidal modulation with a pulsed fraction of ~7 %. In it, four phase intervals can be easily identified (see figure 2).

Figure 3 shows the four spectra (colour coded) corresponding to the four phase intervals. The bulk of the phase dependence of the spectra is seen to be due to the absorption lines variation, while the continua, as seen in the spectral regions not affected by the lines, remain reasonably constant. Indeed, one could go as far as saying that the X-ray source pulsation is largely due to the phase variation of the lines with the neutron star rotation. It is the first time that such a phenomenon is observed for a rotating compact object.

Four colour-coded stripes in the figure inset allow for a direct appreciation of the absorption feature phase variation, as seen through the spectral residuals. To summarise, the soft X-ray spectrum of 1E1207.4-5209 is characterised by three absorption features with integrally spaced energies (0.7, 1.4 and 2.1 keV; there may even be a fourth one at ~2.8 keV) and which are varying in phase with the neutron star rotation, possibly accounting for the bulk of the X-ray pulsation. They also have, in our deep XMM-Newton observation, an extent, or depth, which appears to be globally decreasing with energy (see figure 1), although this effect appears to be phase-dependent.

X-ray spectral features from an Isolated Neutron Star (INS) can be due to atomic transition lines, photoabsorption edges or cyclotron lines.

The atmosphere of an INS is a very difficult environment for physics, owing to the necessity of working with an a-priori unknown nuclear species composition, and level of ionisation, in a Landau regime of high magnetic field. We refer to the extensive work of [17,18] on possible transition lines or photoabsorption edges. Their final preference for explaining the Chandra features in 1E1207.4-5209 with He-like mid Z (O or Ne) absorption lines, although certainly self-consistent, seems overtaken by the present XMM-Newton data.

Atmospheric absorption models [19,1] also meet with the strong difficulty of the lack of similar features in any of the several INSs for which good Chandra/XMM-Newton spectra are now available. Data for such diverse INSs as Vela, PSR B0656+14, PSR B1055-52, the msec pulsar PSR J0437-4715 and the two "dim" INSs RX J1856.5-3754 and RX J0720.4-3125 (see[5,6] for recent reviews), were explored and none shows absorption or emission features. These objects cover a vast range in age (from $10^4$ y to "eternity" in the case of the msec pulsar), in B fields, in possible interstellar accretion rate, etc . It is difficult to imagine none of these bona fide INSs showing a phenomenology similar to 1E1207.4-5209, if the features of the latter were due to atmospheric absorption.

Occam's razor, in the end, points clearly to cyclotron resonance scattering for explaining the features presented above: three lines, correctly spaced by integral factors (within the uncertainties on their measured centroids) and with a phase variation naturally following the pulsar B-field rotation, as observed for the much brighter accreting neutron stars in binary systems, e.g. [20,21]. Note that small deviations from the 1:2:3 ratio of the line centroids can also de induced by the complex geometrical and physical environment related to harmonic generation [22]. Our best fitting temperature for the source continuum emission ( about 0.21 keV) is nicely consistent with the equilibrium Compton temperature due to resonant scattering, computed by [2] to be .27 of the energy of the first harmonic, or about 0.19 keV.

Following Freeman et al. [23], we note that in cyclotron scattering above the surface of an INS, the cross section (and the relative strength of the harmonic lines) depends on the angle between the observer and the magnetic field. This provides a natural explanation for the phase variation of the feature *depths* . There are also compelling physical reasons [23,24] why all line *widths* should vary in phase, with maximum broadening appearing when the observer's line of sight and the pulsar B field are perpendicular. Of course, such asymmetric and varying line profiles render problematic any precise evaluation of each line's central energy as well as their harmonic spacings. Thus, a fortiori, considerations on line shapes and their centroid energies based on the phase-averaged spectra of Figure 1 have little physical meaning. Rather, any physical interpretative work on 1E1207.4-5209 should be driven by phase-resolved spectroscopy.

If the ~0.7 keV line is the fundamental electron cyclotron frequency, then B averaged over the star's disc is ~6 $10^{10}$(1+z), or ~8 $10^{10}$ G, assuming a "standard" 20% gravitational redshift. For protons, the field should be a factor $m_p/m_e$ higher , ~ 1.6 $10^{14}$ G, i.e. a magnetar type magnetic field [25]. Neither of these values agrees with the B field inferred from the 1E1207.4-5209 timing parameters, which give 2-3 $10^{12}$G under the rotating dipole hypothesis. The latter, however, is not problem-free for our object since the current Pdot = (1.4 ± 0.3) $10^{-14}$ s/s implies a pulsar age of ~4.8 $10^5$ y, totally incompatible with that of the supernova remnant associated to it, no more than ~ $10^4$ y old [26]. While the reported tentative radio detection [27] may help to better monitor the pulsar timing parameters, it seems unlikely that better timing could solve the age inconsistency problem. More drastic assumptions, such as low spin frequency at birth, should possibly be considered.

To reconcile the pulsar period derivative and the B value inferred in the case of electron cyclotron emission close to the neutron star surface, one should invoke additional pulsar braking mechanisms, such as the debris disk configuration proposed e.g, by [28] . Of course, electron cyclotron scattering at R~3-4 $R_o$ (i.e. 20-30 km above the neutron star surface) would fit all the observations.

Alternatively, the cyclotron features could be due to protons. This would require an improbable surface magnetic field ~100 times stronger than that deduced from the spindown. Although a single proton cyclotron feature could be present in the optical spectrum of Geminga[29], the B-field requirements for 1E1207.4-5209 seem too large to support the proton cyclotron explanation for the observed X-ray features.

On the whole, electron resonance scattering in an abnormally low magnetic field would seem to best explain the current XMM-Newton observations. The cyclotron nature of the features, as opposed to the atmospheric one, would also naturally explain the uniqueness of the 1E1207.4-5209 spectrum in the INS context, since the powerful combination of throughput and energy resolution of XMM-Newton/EPIC is only effective for less than a decade in photon energy. Most INSs, with a more normal B-field, may have electron cyclotron features in the 10-20 keV range, i.e. where such features were observed for the much brighter accreting objects.

**Acknowledgements**

It is a pleasure to thank the XMM-Newton Project Scientist Fred Jansen and all the XMM-Newton team in Vilspa for approving and carrying out this target of opportunity observation on a very short notice. We are also grateful to W.Becker and M. Turner for helping us to optimise EPIC settings. GFB gladly acknowledges enlightening discussions with Don Lamb, George Ricker and John Doty. The XMM-Newton data analysis is supported by the Italian Space Agency (ASI).

ADL acknowledges an ASI fellowship.

**Figure captions**

Figure 1

*Spectra collected by the pn (Figure **1a**) and MOS (**1b**) cameras on the EPIC instrument during the observation of Aug 4-6 2002.* While the two MOS cameras were operated in "full frame" mode [13], the pn one was in "small window" mode for accurate timing of source photons (6 ms resolution, ref.12). All cameras used the thin filter[13]. The data were processed with the XMM Newton Science Analysis Software, version 5.3.3. After removing time intervals with high particle background and correcting for dead time, we obtain a net exposure of 138.4 ksec for the pn camera and 193.6 and 194.7 ksec for the mos1 and mos2, respectively.

Our count rates, 1.486 ±0.003 c/sec for the pn and 0.379 ±0.001 c/sec (mos1) 0.388 ±0.001 c/sec (mos2), are slightly larger than those reported by [11], owing to our use of the thin filter.

Data points and best fitting continuum spectral models are shown, together with residuals in units of standard deviations from the best fitting continuum. The best spectral model ($\chi^2_\nu$=1.25, 111 d.o.f.) describes the continuum as the sum of two black body curves with kT= 0.211 ± 0.001 keV, for an emitting radius R=2.95 ± 0.05 km (assuming a distance d=2.2 kpc[26]), and 0.40 ± 0.02 keV (R=250 ±50 m). $N_H$ = (1.0 ± 0.1) $10^{21}$ cm$^{-2}$, to be compared with 1.6 $10^{21}$ cm$^{-2}$ estimated from radio observations of the associated supernova remnant G296.5+10.0 [26]. Any other model for the continuum fails to give such a good fit to the data: we tried a composite (black-body+power law) model ($\chi^2_\nu$=1.9, 111 d.o.f.) and a Hydrogen atmosphere model of [19] ($\chi^2_\nu$=1.7, 113 d.o.f.). The central energies of the three absorption features are 0.72±0.02 keV, 1.37±0.02 keV and 2.11 ±0.03 keV. The possible fourth feature in the pn spectrum is at 2.85±0.06 keV. The optical depths τ for the two main features are similar ($\tau_1$=0.54±0.02; $\tau_2$=0.52±0.02); however, the integrated number of scattered photons is more than 4 times higher for the feature at 0.7 keV. The third feature is less intense. All errors are at the 90% confidence level.

Figure 2

*Total light curve relative to 208.000 photons collected by the pn camera folded at the best period of P= 0.42413076 s* . When compared to the Chandra measurement of January 2000 [9,30], this yields a refined estimate of the pulsar spindown of Pdot = (1.4 ± 0.3) $10^{-14}$ s/s. The four equal phase intervals used to assess the spectral variation (namely, "peak", "decline", "trough" and "rise") are also shown.

Figure 3

*Comparison of the four pn spectra for phase intervals as defined in figure 2*. Note that absolute (counts/sec/keV) spectra are plotted.

Colour code: black "peak", red "decline", green "trough" , blue "rise".

The peak of the total light curve corresponds to the phase interval where the absorption lines are at their minimum (black points) while the light curve trough happens when the absorption lines are more important (green points).

Inset : The four panels show the residuals of the phase dependent spectra from the two-black-body fit used in figure 1a, colour coded as before. Phase variations in both shape and intensity for all features are apparent. The comparison of phase dependent residuals with figure 3 of [11] shows the impressive improvement achieved with the current long observation.

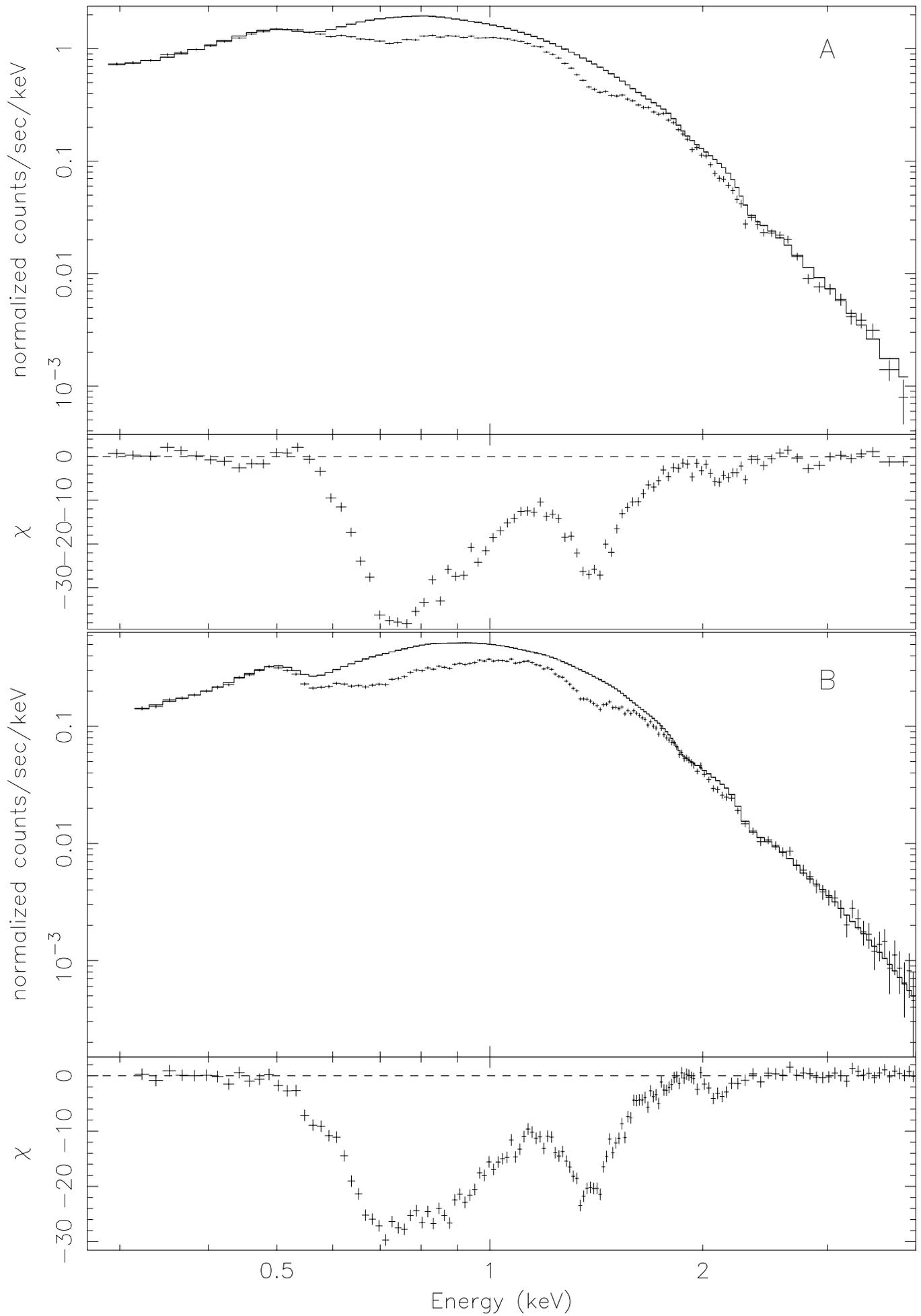

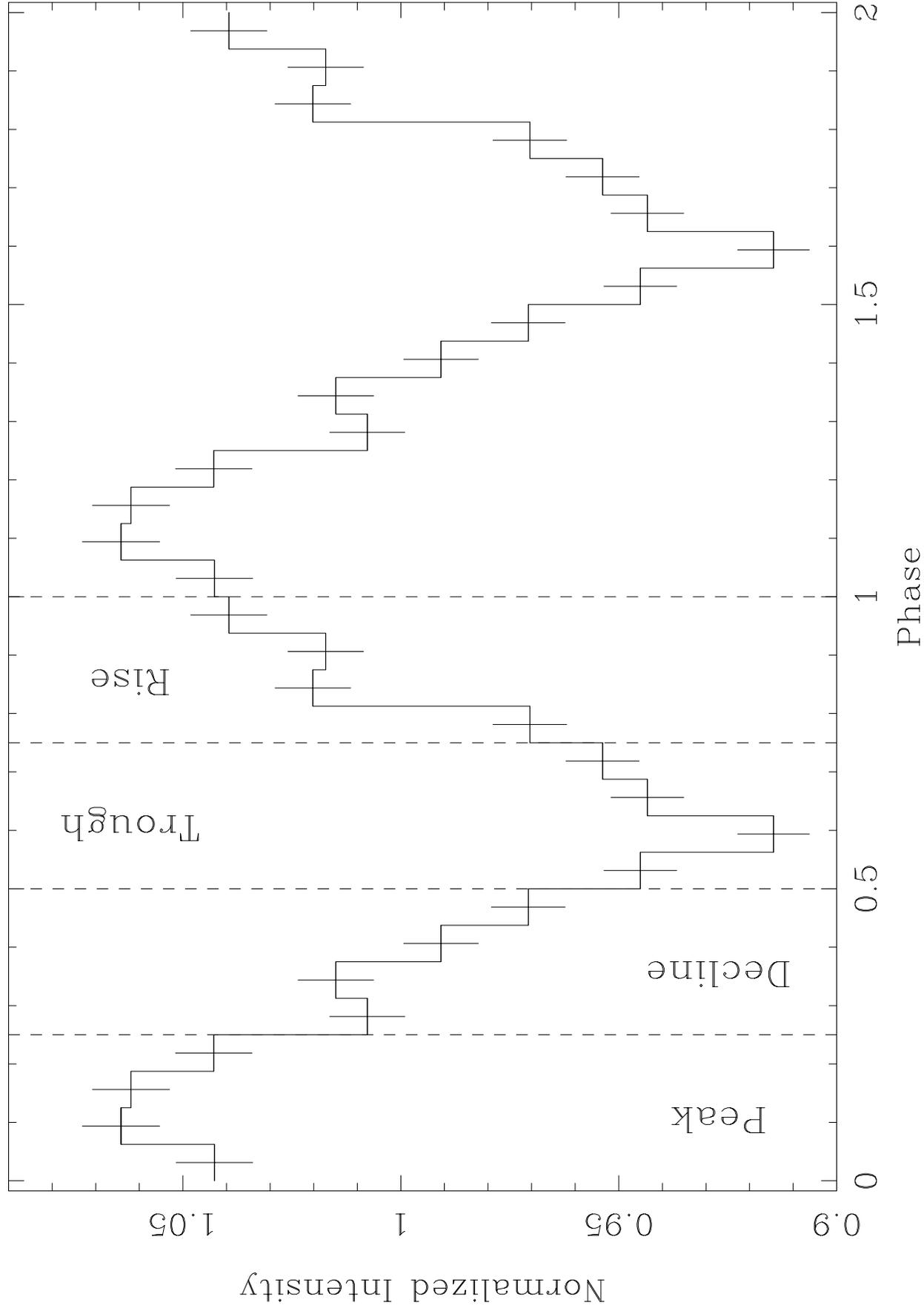

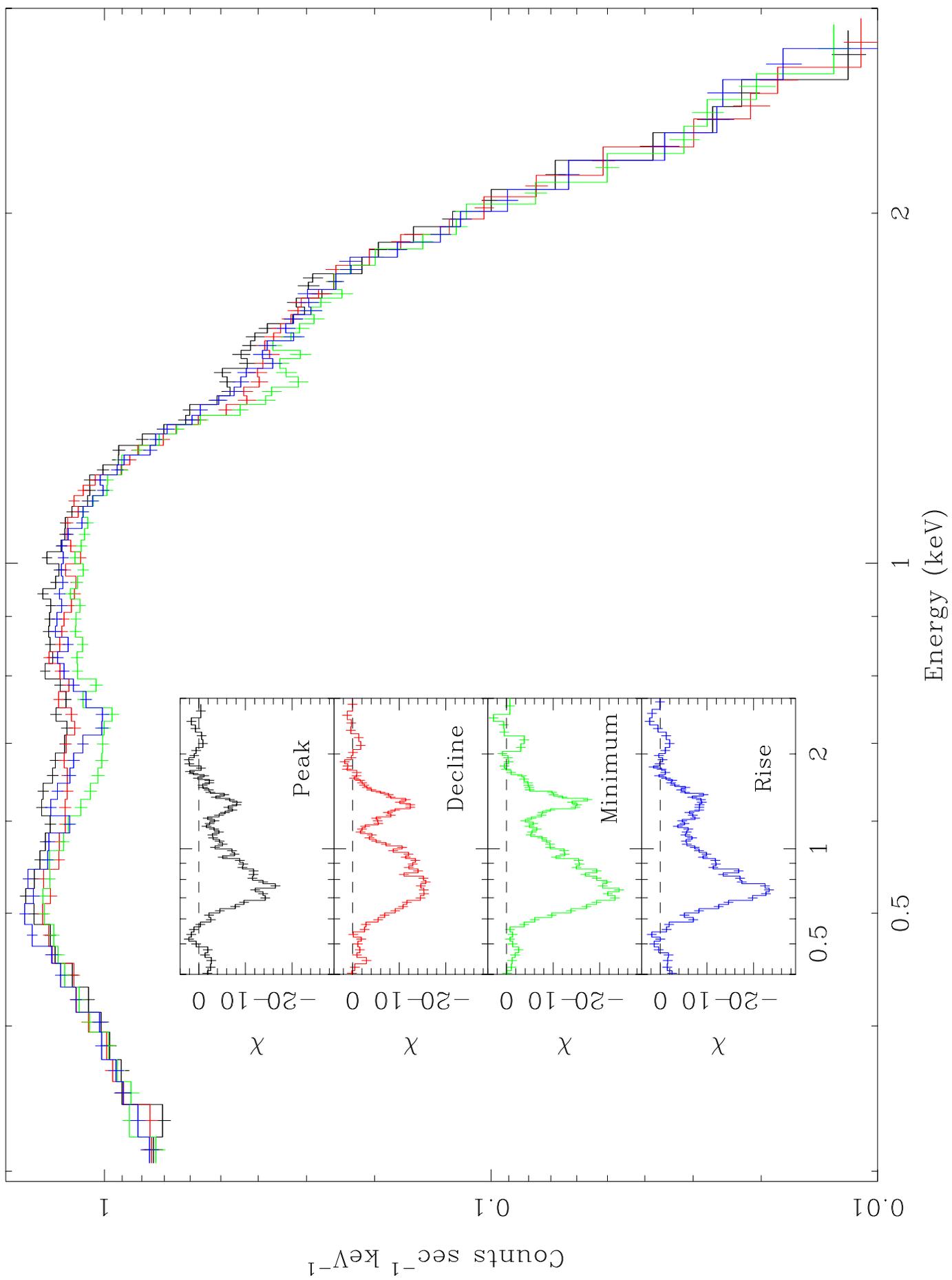